# Developing Potential Energy Surfaces for Graphene-based 2D-3D Interfaces from Modified High Dimensional Neural Networks for Applications in Energy Storage

Vidushi Sharma*, Dibakar Datta*,

Department of Mechanical and Industrial Engineering, New Jersey Institute of Technology, Newark, NJ 07103, USA

## Abstract

Designing a new heterostructure electrode has many challenges associated with interface engineering. Demanding simulation resources and lack of heterostructure databases continue to be a barrier to understanding the chemistry and mechanics of complex interfaces using simulations. Mixed-dimensional heterostructures composed of two-dimensional (2D) and three-dimensional (3D) materials are undisputed next-generation materials for engineered devices due to their changeable properties. The present work computationally investigates the interface between 2D graphene and 3D tin (Sn) systems with density functional theory (DFT) method. It uses computationally demanding simulation data to develop machine learning (ML) based potential energy surfaces (PES). The approach to developing PES for complex interface systems in the light of limited data and transferability of such models has been discussed. To develop PES for graphene-tin interface systems, high dimensional neural networks (HDNN) are used that rely on atom-centered symmetry function to represent structural information. HDNN are modified to train on the total energies of the interface system rather than atomic energies. The performance of modified HDNN trained on 5789 interface structures of graphene|Sn is tested on new interfaces of the same material pair with varying levels of structural deviations from the training dataset. Root mean squared error (RMSE) for test interfaces fall in the range of $0.01 - 0.45$ eV/atom, depending on the structural deviations from the reference training dataset. By avoiding incorrect decomposition of total energy into atomic energies, modified HDNN model is shown to obtain higher accuracy and transferability despite limited dataset. Improved accuracy in ML-based modeling approach promises cost-effective means of designing interfaces in heterostructure energy storage systems with higher cycle life and stability.

## 1. INTRODUCTION

Intricately engineered device designs are put forth very often in recent years under the broad spectrum of advanced technology to target sustainability and enhanced efficiency. Complexity revolving around such device designs calls for the discovery of new multicomponent systems or new materials altogether. The concept of engineering materials has been applied in the vast sense in multicomponent systems to enhance functionality based on structure-function relationship [1]. For instance, in energy storage alone, the development of electrode design as core-shell, layered, or coating is continuously practiced for enhancing the electrode capacity and cycle life [2-6]. Such arrangements of different materials together in engineered devices create unique interfaces whose equilibrium and properties remain of intense research interest.

Two-dimensional (2D) material-based heterostructures (2D + nD, n=0,1,2,3) have attracted enormous interest in various fields [7-9], including batteries [10, 11]. The discovery of graphene in 2004 [12] opened a new space for mixed-dimensional materials where heterostructures based on graphene-like 2D materials layered with three-dimensional (3D) bulk materials have exhibited intriguing properties [7, 13]. The mechanical, electrical, optical, magnetic, and thermal properties exhibited by these heterostructures have applications in batteries, optics, solar technologies, and electronics [14-22].

Arrangement of 2D-3D materials can be achieved either during epitaxial growth of one material over the other or with post-growth modification techniques [13]. Despite the development of multiple synthesis modes, a complete characterization of these complex nanoscale interfaces is yet a challenge. Scanning tunneling microscopy (STM) and atomic force microscopy (AFM) has enabled studying the interface between ordered 2D material heterostructures [23]. Yet, their scope is not expandable to polymorphing interfaces where 3D bulk undergoes lattice distortions and phase transitions to stabilize itself over a 2D substrate.

The fundamental understanding of interfacial properties in these systems is critical for sustaining the desired electro-chemo-mechanical behavior. Quantitative computational modeling efforts with ab initio methods such as density functional theory (DFT) indicate the influence of structural polymorphism in 3D bulk on interfacial properties such as interfacial strength, electron transfer, and conductivity [24, 25]. However, studies in this domain remain less explored due to the high computational cost involved in ab initio methods. Moreover, molecular simulations of interfacial properties need reliable inter-atomic potential, limited to only a few elements. Due to these restrictions, there has been no significant advancement in interface studies at the atomic/molecular level. These limitations are expected to be overcome with the emergence of artificial intelligence (AI) based models in materials.

Machine learning (ML) based potential energy surface (PES) can describe complex systems at low computational cost and with close to first principles accuracy. These methods rely on a large amount of DFT data of materials (structures, energies, and forces) to efficiently explore the chemical space with respect to the target properties during training. Atomic structures of materials are represented by appropriate descriptors and fed to the neural network (NN) algorithm to generate PES that is invariant to translational, rotational, and permutation of homonuclear atoms [26, 27]. Such PES are independent of any physical parameters and approximations, unlike empirical force fields, and hence are more accurate if descriptors describe the atomic local environments efficiently. Several ML techniques have been employed for PES development: linear regression [28, 29], Gaussian approximation [30, 31], high-dimensional neural networks (HDNN) [32] and graph neural networks (GNN) [33, 34]. Of particular interest are HDNN using atom-centered symmetry functions (ACSF) as input descriptors [26] that have been successful for a wide range of materials due to their generality [35-38]. Consequently, a lot of effort has been spent on refining the original HDNN given by Behler and Parrinello[32] to develop more efficient approaches [39-41].

The development of ML-based PES is still in its nascent stage and has barely explored materials' domain beyond small organic molecules [42-44], single component condensed systems [32, 45, 46], bicomponent systems [47-49], and crystalline ordered motifs [50]. ML research in heterostructure territory is mostly limited to searching and predicting new ordered heterostructures with targeted properties [51-53]. The recent attempts to model 3D-3D interface grain boundaries demonstrate the scope of ML models in capturing mechanics at nanoscale when trained on high accuracy data [54]. However, the primary bottleneck ML needs to scale in modeling complex materials is the tradeoff between quantity and quality of training data. As the complexity of the problem increases, acquiring ample data either computationally or empirically becomes a challenge.

The present work explores a mixed dimensional heterostructure (2D-3D) interface formed by graphene (Gr) and tin (Sn) for the lattice distortions by utilizing DFT. Geometry optimization of Sn bulk over Gr results in the lattice rearrangements in interfacial Sn atomic layers. The extensive computation undertaken to study these interfaces with DFT has been further used to develop an ML-based PES for Gr and Sn interfaces. Here, the development of ML models could reinforce the atomic realization of these complex interfaces. This is one of the first attempts at modeling complex polymorphing 2D-3D interfaces with ML. We use a modified approach to HDNN algorithm to develop PES for Gr|Sn interfaces which exhibit good transferability to new Gr|Sn interface systems despite limited training data. The design and utilization of ML-based potentials can be extended to expedite interface simulations in the future. Their scope of utilization in overcoming existing battery electrode design issues has been presented as a comprehensive discussion in the last section.

## 2. METHODOLOGY AND COMPUTATIONAL DETAILS

### 2.1 High Dimensional Neural Networks (HDNN)

The present work uses second-generation high dimensional neural networks (HDNN) generalized by Behler and Parrinello in 2007 [32]. Limitations possessed by first-generation feedforward neural networks in addressing the full dimensionality of large-condensed phase systems are overcome by assuming that the total energy of the system $E_{total}$ can be disseminated into energy contributions by each individual atom ($E_i$) based on their local chemical environment.

$$E_{total} = \sum_{i=1}^{N} E_i \tag{1}$$

A large part of this assumption relies on the accuracy in describing atomic interactions in the local chemical neighborhoods. Appropriate descriptors that can convert cartesian coordinates to numeric vectors describing atomic interactions with translational, rotational, and permutational invariance are critical for the PES development. The descriptor defines the chemical environment of each atom within a specific cutoff radius ($R_c$). A large $R_c$ value ensures that all energetically relevant interactions such as covalent, electrostatic, dispersion, and van der Waals interactions are included. Once the descriptor obtains fingerprints of atomic local space, separate feedforward neural networks are used for each atom in the system to express atomic energy contributions $E_i$. These $E_i$ are then added to yield the total energy of the system $E_{total}$.

HDNN model in the present work (Figure 1b) is a modified version of second-generation HDNN for bicomponent systems (also called BPNN for Behler and Parrinello Neural Networks, Figure 1a). Figure 1 differentiates both HDNN models for a system of six atoms, three Sn and three C. In the BPNN model, chemical species are separated by a distinct set of weights. Atomic input features defined by descriptors ($G_i$) are fed to atomic neural networks (*ann*), and two sets of *ann* weights are fitted corresponding to each chemical species in the system. Weights corresponding to Sn atoms (*set-a, red-ann*) are identical, as are the weights corresponding to C atoms (*set-b, yellow-ann*). This ensures the invariance of total energy against interchanging of two identical atoms within set-a and set-b. In addition, this permits easy size extrapolation of the model. If another atom is added to the system, additional *ann* corresponding to the species can be appended to the architecture and added to the total energy expression (represented in equation 1). Complete details of second-generation HDNN (BPNN) for multicomponent systems can be found in [55]. In contrast, modified HDNN in Figure 1b identifies chemical species with an added input feature rather than relying on separate trained weights. Thus, weights of all *ann* (*set-k*) are trained to be identical for the Sn-C system. All *ann* architectures and weights are suited for a bicomponent system of Sn-C rather than an individual species. This approach permits easy system size extrapolation.

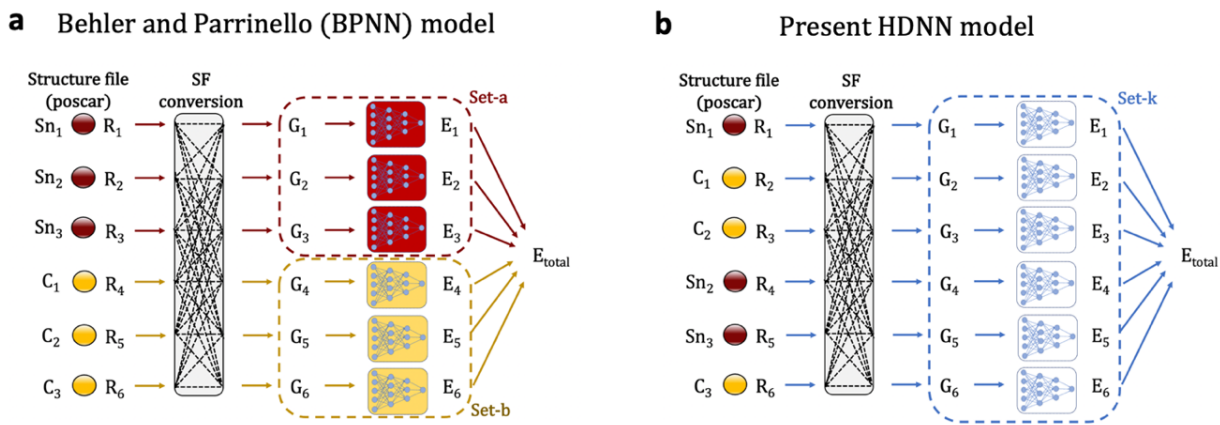

**Figure 1 :** Comparative schematics of high dimensional neural networks (HDNN) for bicomponent (Sn|C) system. (a) HDNN by Behler and Parrinello (BPNN) for bicomponent systems where weights and architecture of atomic neural networks(*ann*) are same for single chemical species. Red-*ann* in set-a corresponds to Sn atoms and yellow-*ann* in set-b corresponds to C atoms. (b) Modified HDNN in present study for bicomponent systems. Weights and architecture of all atomic neural networks (*ann*) are same and correspond to the Sn|C system rather than single species. Atomic species are differentiated by added feature of atomic number.

## 2.2 Atom-Centered Symmetry Functions (ACSF or SF)

There have been several attempts to use cartesian coordinates as structural inputs for ML-based PES [39, 56]. These are recognized to be a poor choice. Cartesian coordinates are not independent of molecular translation and rotation. Since neural networks are mathematical fitting methods, the output is sensitive to absolute values of input features. To overcome these limitations in describing complex chemical structures to HDNN, Behler et al.[32] introduced atom-centered symmetry functions (ACSF) that describe the chemical neighborhood of each atom with the help of radial and angular symmetry functions. There are two types of ACSF commonly used that define radial($\mathbf{G^2}$) and angular($\mathbf{G^4}$) information of the central atom's neighborhood within the sphere defined by the cutoff radius ($\mathbf{R_c}$). The cutoff function for all the neighboring atoms within $\mathbf{R_c}$ is defined as

$$f_c\left(R_{ij}\right) = 0.5 \times \left[\cos\left(\frac{\pi\, R_{ij}}{R_c}\right) + 1\right] \tag{2}$$

Where $\mathbf{R_{ij}}$ is the distance between central atom $\boldsymbol{i}$ and its neighboring atom $\boldsymbol{j}$. $\boldsymbol{f_c(R_{ij})}$ is a continuous and differentiable function whose value turns 0 when $R_{ij} > R_c$. The radial and angular SF for central atom $\boldsymbol{i}$ are defined with the help of the cutoff function as two-body and three-body sums.

$$G_i^2 = \sum_j e^{-\eta(R_{ij}-R_s)^2} \cdot f_c\left(R_{ij}\right) \tag{3}$$

$$G_i^4 = 2^{1-\zeta} \sum_{j,k \neq i}^{all} \left(1 + \lambda \cos\theta_{ijk}\right)^{\zeta} \cdot e^{-\eta\left(R_{ij}^2 + R_{ik}^2 + R_{jk}^2\right)} \cdot f_c(R_{ij}) \cdot f_c(R_{ik}) \cdot f_c(R_{jk}) \tag{4}$$

Here, $\mathbf{G_i^2}$ is a sum of Gaussians multiplied by the cutoff function. The width of the Gaussian and the center of the Gaussian can be defined by parameters $\boldsymbol{\eta}$ and $\mathbf{R_s}$. A non-zero $\mathbf{R_s}$ value can shift the center of Gaussian away from the reference atom. Therefore, it is preferably set to 0. The parameter $\boldsymbol{\eta}$ is a Gaussian exponent responsible for indicating reduced interaction strength with increasing distance between the two atoms. Parameters $\boldsymbol{\zeta}$ and $\boldsymbol{\lambda}$ in the function $\mathbf{G_i^4}$ control the angular resolution and cosine function, respectively. $\boldsymbol{\lambda}$ usually takes value either +1 or -1 for inverting the cosine function maxima from $\theta_{ijk} = 0°$ to $\theta_{ijk} = 180°$ [26]. The most preferred value for $\boldsymbol{\zeta}$ is 1 as it provides sufficient coverage centered at 0° (when $\lambda = 1$). Higher values can increase angular resolution close to the center at reduced coverage cost [57]. Multiple parameter-set for each SF type are needed to cover different portions of chemical environment. Values of these parameters define the high dimensional input vector representing the local environment of each atom in the material system. It is advisable that 100-150 $\mathbf{G_i}$ be used for the bicomponent system, such as the interface systems, to capture full dimensionality of the system. Redundancy of the information that can arise due to large number of $G_i$ has been recognized to be not a problem for HDNN [26]. SF are uniquely beneficial as input vectors for HDNN because input vector size is independent of the actual number of neighboring atoms within a set cutoff radius $R_c$ [36]. We used DScribe package in Python to calculate SF for interface systems with the parameters defined in Table 1 [58].

We use a range of $\eta$ values to capture the full dimensionality of the structures. The presented parameter set is chosen based on the benchmarking studies on descriptors for bicomponent bulk systems having elemental makeup similar to our structures [35, 59, 60]. $\lambda$ has both values +1 and -1 corresponding to both centers in the SF set. To attain high angular resolution as well as complete coverage for intended interface systems, we use higher values for $\zeta$ in addition to 1 ($\zeta$ = 1, 2, 4). These SF set yielded the best results in our comparative evaluation and served as the foundation for further assessment of our model. As described in section 2.1, we preferred to use large $R_c$ for sufficient atomic interaction coverage. Here, $R_c$ = 8.9Å is found to be sufficient for optimum coverage of atom's local environment (validation presented in supplemental information section I). By using the SF parameter set in Table 1, the chemical environment of each atom was represented by 162 input features. It is important to note here that DScribe package used for conversion of cartesian coordinates to SF considers atomic number ($Z$) of chemical species in the environment of central atom by appending the value to the SF ($G^2$ and $G^4$). However, it does not consider atomic number of central species in any way. To overcome this drawback, an additional input feature was added to the SF input feature for each atom which was its own atomic number. Thus, each atom was represented by 163 input features in our study.

**Table 1**  Parameters Used to Compute the ACSF in the Study

| Descriptors | Parameters | Values |
|---|---|---|
| $G^2$ | $R_c$ (Å) | 8.9 |
| | $R_s$ (Å) | 0 |
| | $\eta$ (Å$^{-1}$) | 0.003214, 0.035711, 0.071421, 0.124987, 0.214264, 0.357106, 0.714213, 1.428426 |
| $G^4$ | $R_c$ (Å) | 8.9 |
| | $\lambda$ (Å) | -1,1 |
| | $\zeta$ | 1, 2, 4 |
| | $\eta$ (Å$^{-1}$) | 0.003214, 0.035711, 0.071421, 0.124987, 0.214264, 0.357106, 0.714213, 1.428426 |

Note:  Several values of $R_c$ were tested and the value of 8.9Å was found to give optimum results for presented 2D|3D interface systems.

## 2.3 Density Functional Theory (DFT) Computation Details

Coherent interface models were created between ordered single layer Gr and Sn allotropes with an optimum interfacial gap of 3.5 Å. Gr contains 60 $sp^2$ hybridized carbon atoms in all the interface structures, whereas the size of Sn bulk varies from atom size of 16 to 64. Sn is known for having many energetically similar allotropes, with alpha ($\alpha$-Sn) and beta ($\beta$-Sn) being the prominent ones. The interfaces are set in the periodic x-y plane across Sn (100) miller indices. These interface structures were modeled with a vacuum of 15Å in z dimensions to circumvent the periodic influences, followed by DFT optimization to obtain relaxed strain-free interface configurations. All crystalline Sn bulks were derived from the materials project database [61]. Amorphous Sn was created with computational quenching of $\alpha$-$Sn_{64}$ following the methodology discussed in our earlier works [24, 62, 63]. Both Gr and Sn structures were DFT optimized individually before interfacing. All DFT calculations are done using VASP [64]. Projector-augmented-wave (PAW) potentials are used to mimic inert core electrons, while the valence electrons are represented by plane-wave basis set [65, 66]. Plane-wave energy cut-off and convergence tolerance for all relaxations are 550 eV and $10^{-6}$ eV, respectively. The GGA with the PBE exchange-correlation function are taken into account [67] with inclusivity of vdW correction to incorporate the effect of weak long-range van der Waals (vdW) forces [68]. The energy minimizations are done by conjugate gradient method with Hellmann-Feynman forces less than 0.02 eV/Å. Considering the vacuum slab structure of all interfaces, gamma-centered k-meshes 4 X 4 X 1 are used for good convergence.

## 2.4 Training and Testing Dataset

Data for training and testing are systematically selected to meet the primary objective of the study, which is developing PES for Gr|Sn interfaces with the least possible computation necessary and best possible transferability. While material databases are a reliable source for most material's structural data for learning-based workflows, they are significantly lacking in the domain of interfacial configurations. Tracing the equilibration of interfaces where exists the possibility of prominent lattice distortions, proved to be computationally expensive. It took approximately 700 hrs with 72 CPU cores to finish the complete relaxation of a single Gr|$\alpha$-Sn interface system. To

initiate 2D-3D interface-based structural analysis in the future, there is a need to develop machine learning-centered cheaper and faster workflows.

We optimized multiple unequilibrated Sn and Gr interface structures and divided them into training and testing datasets. The training dataset consists of five Gr|Sn interfaces: crystalline $\alpha$-$Sn_{(32\ and\ 64)}$ , $\beta$-$Sn_{(16\ and\ 18)}$, and amorphous $Sn_{64}$ interfaced with Gr (shown in supplemental information section II). The training dataset is built from convergence iterations of DFT simulations which covers the trajectory of minima search for five interfaces starting from the initial non-equilibrated structure. This scheme ensured that non-equilibrated and intermittent interface structures were as much part of the learning process as the relaxed structures. The intermediate DFT iterations of five Gr|Sn interface structures accounted for 5789 structures with their reference energies for training.

To analyze the performance of our model and test the transferability of PES in a sequential order of unfamiliarity, we use four carefully contemplated test interface structures (Figure 2). The first test structure (T1) is the very Gr|$\beta$-$Sn_{16}$ interface used in the training dataset, except that the orientation of Sn bulk is slightly shifted over Gr surface in T1. It is an example of known interface structure and unknown orientation. The second test structure (T2) is again a familiar interface with increased Sn bulk size (Gr|$\beta$-$Sn_{16}$ $\rightarrow$ Gr|$\beta$-$Sn_{32}$). The objective of T2 is to test the system size extrapolation capabilities of the PES (see figure S2 in supporting information section II).

In contrast to T1 and T2, the third test structure (T3) is completely unfamiliar interface. A new Sn bulk (mp-949028 from Materials Project Database) is interfaced with Gr. The fourth and final test interface (T4) is derived from T3 by creating divacancy defects in the interfacing Gr. This change adds complexities of defects and surface adsorption in the interface structure, which are not noted in the earlier test interfaces. Differences in test structures from the training dataset are summarized in Table 2. The initial test interface structures were created like training interfaces and subjected to DFT optimizations. Since the current machine learning scheme does not include automated equilibration, the ability of PES to predict energies of intermediate configurations as structures search for global minima is assessed by predicting energies of test interfaces between initial and final configurations. The variations in the test interfaces during DFT optimization can be noted in the iteration snapshots presented in supplemental information section III. While minimal Sn alignment changes are observed in initial - final T2 and T3 structures, major structural transformative and surface defects are seen in T1 and T4, respectively.

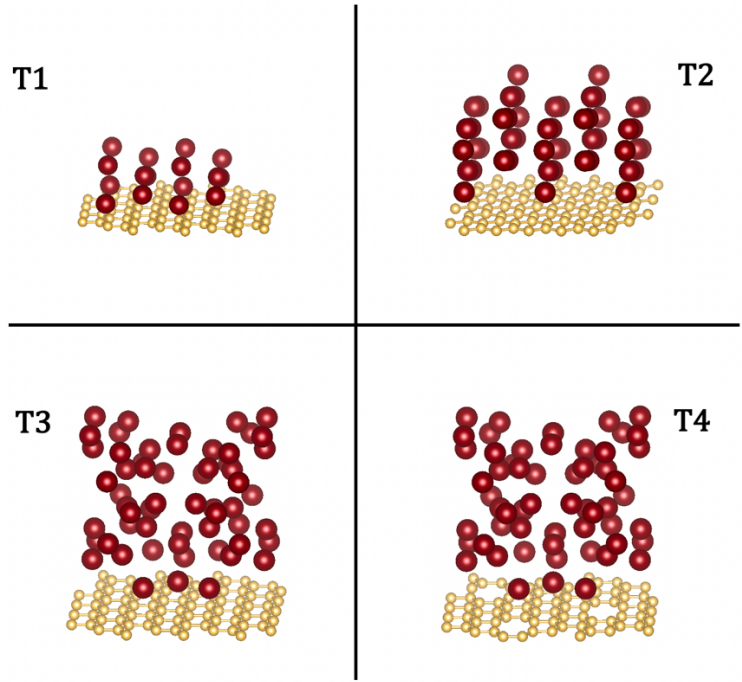

**Figure 2:** Test interface structures (T1-T4) labelled in the order of unfamiliarity.

**Table 2**    Notable Differences in Test Structures from Training Dataset

| Notable features of test structures | T1 | T2 | T3 | T4 |
|---|---|---|---|---|
| Familiar interface | O | O | × | × |
| Familiar interface orientation | × | O | × | × |
| Similar Sn bulk size | O | × | × | × |
| Familiar Sn allotrope bulk | O | O | × | × |
| Familiar Gr substrate | O | O | O | × |

## 3. RESULTS AND DISCUSSION

### 3.1 Phase Changes at Graphene|Sn Interface

This section discusses the atomic specifications of the Gr|Sn interface in optimized structures, which predominantly set apart these interfaces from Gr-based interfaces reported previously [7, 13]. Discussion of interface phenomenon is important for designing interfaces and controlling heterostructure properties in applied technologies. The Gr|Sn interface systems are optimized by DFT to obtain relaxed strain-free interface configurations. Presented interfaces structurally resemble interfaces assembled in devices post-growth rather than interfaces originated during direct epitaxial growth of Gr on a substrate. Final interfacial configurations of Gr|Sn are depicted in Figure 3 for two prominent Sn allotropes, α-Sn, and β-Sn, respectively. α-Sn is a diamond cubic crystal and β-Sn is a body-centered tetragonal crystal (Figure 3b), which are two solid allotropes of Sn commonly in use. At temperatures below room temperature (286 K), α-Sn is the stable phase, which transitions to its β configurations rather quickly as temperature rises [69]. Sensitivity of temperature conditions symbolizes the significance of solid Sn α↔β transitions for practical applications. This sensitivity elevates in the interfacial conditions with large lattice mismatch.

Differences in materials and lattice constants imply strained conditions in the buffer layer of Sn at the Gr|Sn interface, which conditions the Sn bulk towards a possible phase change. Consequently, the observed lattice constant of interfacial Sn ($c = 4.5$Å) is different from the rest of the α-Sn bulk ($c = 4.7$Å) in Figure 3a. This indicates a possible phase transformation from α-Sn to β-Sn in the buffer layer at Gr|α-Sn interface. However, these structural transformations of Sn are at a few layers limit at the surfaces and do not proliferate to central regions of the bulk where α conformations are retained (Figure 4). Surface relaxation of independent α-Sn slab did not show any distortions, which eradicates the possibility of this structural reconstruction at Gr|α-Sn interface as mere surface defects. We repeated the simulation with increased vacuum in z dimensions to ensure structural distortions in the top layer are not due to periodic influences (shown in supplemental information section IV). Our observations indicate that this surface hardening of α-Sn is nucleated due to presence of Gr interface. In contrast to Gr|α-Sn interface,

no structural distortions are noted in the relaxed Gr|β-Sn in Figure 3c, indicating the preference of the Gr interface towards β-Sn.

Structural changes in Sn at the Gr interface are also significantly impacted by Sn bulk size. Figures 3d and 3e exhibit Gr|Sn interfaces with smaller α-Sn and β-Sn bulks. Complete distortion of α-$Sn_{32}$ bulk is noted in Figure 3d with an increased density (see supporting information section V). Likewise, β-$Sn_{16}$ rearranged over Gr surface as a single atomic layer in Figure 3e with near-atomic distances of 3.15Å. Drop in the Sn bulk size causes a reduction in dimensions of Sn bulk and brings all the Sn atoms to the surface, much like 2D materials. These observations in small Sn crystals fall in line with prior experimental evidence of Sn nanocrystals becoming denser over graphene surface [70]. The relative differences in the material surfaces and charge analysis of Gr|Sn interfaces strongly indicate the weak van der Waals forces as the foundation of the formed interfaces. Charge analysis was performed in the said interfaces using Bader charge scripts by Henkelman group [71]. The net electron exchange across the Gr|Sn interface is less than 1e$^{-1}$ for all interfaces denoting negligible covalent interaction.

Sn is a well-known high-capacity anode for LIB and NIB with prominent shortcomings from phase transitions (β↔α) and volumetric strains. The presence of Gr substrate for Sn anode provenly scales down the volume expansion associated mechanical failures [62, 70]. Our analysis suggests it can possibly minimize frequented phase transitions from β↔α due to its preference for β-Sn. With our computational study, we attempt to closely understand the swift structural transformations in Gr|Sn interfaces. However, there is scope for further experimental XRD analysis of Sn crystals interfaced with Gr, which can validate presented observations.

Relamed Graphene|α-Sn Interface            Relaxed Graphene|β-Sn Interface

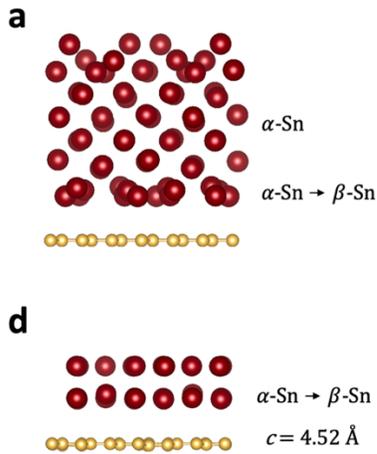
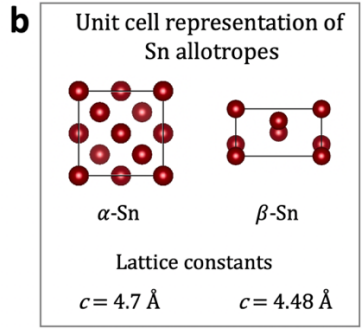
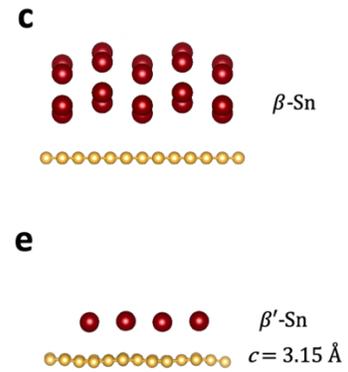

**Figure 3:** DFT relaxed Graphene|Sn interface systems. (a) Side view of relaxed Graphene|α-Sn$_{64}$ interface system. Phase transformations of α-Sn to β-Sn noted in the Sn surface layers due to presence of graphene substrate. (b) Unit cell representations of α-Sn and β-Sn with lattice constants 4.7Å and 4.48Å, derived from materials project database (mp-117 and mp-84) and used for construction of Sn bulks. (c) Side view of relaxed Graphene|β-Sn$_{32}$ interface system. (d) Side view of relaxed Graphene|α-Sn$_{32}$ interface system. Phase transformation of α-Sn to β-Sn noted in the entire Sn bulk with modified lattice constant of 4.52Å. (e) Side view of relaxed Graphene|β-Sn$_{16}$ interface with Sn rearranged over Gr surface as a single atomic layer of modified β'-Sn.

DFT intermediate structures of Graphene|α-Sn interface

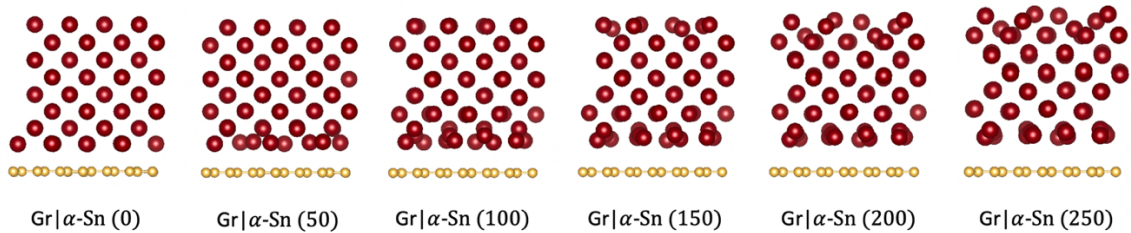

**Figure 4:** Intermediate Graphene|α-Sn$_{64}$ interface structures between initial and equilibrium interface configurations. Structural configuration in first 250 DFT iterations are presented depicting quick structural transformations in early DFT stages. Simulation took approximately 1100 iterations to completely optimize. No major structural rearrangements were noted in the later iterations.

## 3.2 Model Performance

HDNN is traditionally trained on reference energy per atom. However, electronic simulations do not provide actual energy per atom for reference and energy per atom is deduced based on the total energy of the system [26]. Atomic energies are realized by dividing the total energy of the system with the number of atoms. This scheme gives equivalent atomic energies for all the atoms in a system. Such an approximation is suitable for a single component condensed systems where atoms are present in a single phase. However, in section 3.1, it can be observed that Gr|Sn interface structures have multiple phases with strained interfacial Sn atoms. Assuming interfacial Sn atoms will have higher atomic energies than sub-interfacial Sn atoms, training distinct atomic chemical environments in interface systems on uniform atomic energies is not considered a suitable approach.

To validate this assumption, we use two different training approaches: loss calculation with atomic energies ($E_i$) and loss calculation with total system energies ($E_{total}$). In the first approach, model is trained to learn from the reference atomic energies derived from total system energies, as per earlier reports [26]. We use a uniform *ann* architecture throughout the study. Each *ann* consists of 3 hidden layers having 100-50-10 nodes. The input features count is 163 that defines an individual atom's species and chemical environment, as described in section 2.2. Hyperbolic tangent (tanH) activation function is used in the hidden layers, while the output layer giving atomic energy contribution is linear (*ann*: 163-100-50-10-1). Weights and biases are optimized through supervised learning process using Adam optimizer [72] with a learning rate of 0.00001. Loss function after each epoch was determined by the mean squared error of atomic energies from reference DFT energies:

$$Loss = \frac{1}{Size}\left(E_{DFT} - E_{predict}\right)^2 \qquad (5)$$

The batch size for the training is kept 10 (*Size*) and the accuracy of the energy prediction after each epoch is measured in terms of root mean square error (RMSE). Models are trained until accuracy metrics RMSE of at least 0.002 eV/Atom has been achieved. This amounted to 5000 epochs. The performance of the trained model is validated (validation split = 10%), then tested on test structures T1 and T2. The performance of the trained model on 10% validation split results in RMSE 0.0042 eV/Atom. This result is comparable to earlier reported deep learning studies on condensed phase systems [47, 73], thereby concluding that this approach effectively develops PES for interfaces if target interfaces are similar to the training data. Next, trained PES are further used to predict atomic energies of the test structures T1 and T2. The results are summarized in Table 3. While the model performs well on validation split (test data randomly separated out of training data), its performance for new interfaces has been poor, indicating an overfitting case. A potential reason for this performance could be the training process where different atomic environments in a single system are assigned the same atomic energies. This renders the model inefficient in differentiating atomic energies of the atoms with different chemical neighborhoods.

**Table 3**  Performance of PES on Validation Set and Test Interface Structures

| Performance | RMSE in eV/atom |
|:-----------:|:---------------:|
| Validation set | 0.0042 |
| T1 | 0.2235 |
| T2 | 0.9496 |

The second training approach considers the total energy of the system as the final output of modified HDNN. In the last layer of HDNN, atomic energies $\mathbf{E_i}$ from all *ann* are added to yield $\mathbf{E_{total}}$. This required fixing the atomic size (number of atoms in each system) in the training data to be 300. Input features for any system having less than 300 atoms have been extrapolated by zero padding. Inputs ($G_i$) to *anns* for each interface system are shuffled during each epoch. $G_i$ is a one-dimensional array that encodes information about the central atom species and its local chemical environment. Hence, permutations at this stage do not change either the atomic energy or combined total energy. Loss is calculated from the total energies of the system. This allowed model to assign atomic energies based on the chemical space of each atom. Nested *ann* in modified HDNN has 3 hidden layers with 100-50-10 nodes. Hyperparameters of *ann* (*nodes, activation function*) remain the same as described before. Weights and biases are optimized through supervised learning process using Adam optimizer and a learning rate of 0.00001. The loss function after each epoch was determined by the mean squared error of predicted total energies from reference DFT energies as per equation 5. Batch size for the training was kept one. Accuracy of the energy prediction after each epoch is measured in terms of root mean square error (RMSE), and the model is trained on 5789 Gr|Sn interface structures until the total energy RMSE of at least 0.2 eV is achieved.

Once trained, new PES is used to predict energies of test structures obtained from T1 (familiar interface, different orientation). Between unequilibrated and equilibrated T1 structures, there are approximately 260 structural configurations. All of which were used as test data. Figure 5 compares system energies of T1 structures obtained from DFT ($E_{DFT}$) with energies predicted by new PES ($E_{predict}$). Both system energies match closely with the RMSE value 0.0901eV. Slight error is noted for non-equilibrated structures (below 50 DFT iterative structures), which further reduces as the structure stabilizes. Because HDNN are fitted to the total energies of the structure, we note that $E_{predict}$ is bordering on $E_{DFT}$ values but is not equivalent to the exact values even though the T1 interface is very close to trained structural configuration.

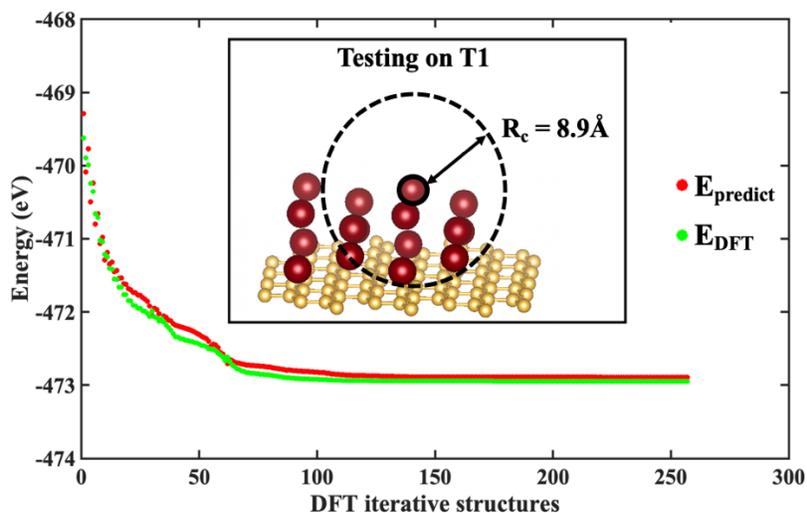

**Figure 5:** Total energies of the test structures T1 predicted by trained HDNN model. $E_{predict}$ and $E_{DFT}$ are total energies predicted by HDNN and DFT, respectively. The dashed sphere with cut off radius Rc =8.9Å represents chemical neighborhood that was observed for all the atoms in the system.

### 3.3 Transferability of PES

The primary objective of the targeted potentials (PES) is the ability to predict energies of new Gr|Sn interfaces and avail the least computation necessary during the development. The HDNN model trained on structures from 5 Gr|Sn interfaces has been tested on new interfaces T1-T4 described in Section 2.4. The results are summarized in Table 4. Between unequilibrated and equilibrated system configurations, there are approximately 260-400 structural configurations for each test structure used for testing. Predicted energies of new interfaces by modified HDNN weights have smaller RMSE values (eV/Atom). RMSE values for T1 and T2 in Table 4 are significantly lower than RMSE values noted in Table 3. This clearly indicates that the deep neural network model designed for such multiphasic interfaces should train across total energies rather than atomic energies to gather complete system information. The difference noted in energies of completely unfamiliar interfaces T3 and T4 is also relatively small. In the absence of accurate empirical potentials in literature, the developed PES demonstrates acceptable performance for new Gr|Sn interfaces constituting structural distortions.

**Table 4**   Performance of PES on Unfamiliar Test Interfaces

| Test interface | RMSE eV/atom |
|:---:|:---:|
| **T1** | 0.016 |
| **T2** | 0.222 |
| **T3** | 0.360 |
| **T4** | 0.458 |

### 3.4 Discussion

Mathematical models like machine learning or deep learning can have powerful predictive accuracy when trained on large datasets. Data requirements are major barriers that delay adoption of ML/DL techniques in artificial intelligence-assisted material development and discovery. The lack of sizable dataset can be compensated with advanced sampling techniques such as heterogeneous dataset, random sampling, and stochastic surface walking [35, 74]. Each of these methods concentrates on a different aspect of PES development. In this work, we use the geometry optimization arc of interfaces as dataset, which includes unequilibrated, metastable, and equilibrated structures. The advisable course of using ab-initio molecular dynamics (AIMD) simulations to explore new equilibria for interfaces prior to DFTs is skipped due to computational demand of the undertaking. The time to run a single AIMD simulation with 500 steps and 2fs timestep on Gr|Sn interface ranged between 400-700 hours on CPU with 72 cores during our initial tests. Despite the limited dataset, our model predicted energies of new interface structures with lower RMSE when compared with the traditional approach. This can be said especially for the interface structures such as T1 and T2, where the Sn phase is familiar with reference training data. By avoiding incorrect decomposition of total energy into atomic energies, the model can obtain a high degree of accuracy and transferability even with a limited dataset.

To overcome the limitations of data, a heterogenous dataset was created consisting of individual Gr and Sn structures and distribution of Sn clusters adsorbed on the surface of Gr (2D-1D interface). This new heterogeneous dataset consisted of 9646 structures (see supporting information Figure S6). However, the modified HDNN model trained on heterogeneous datasets performed poorly for test interface structures (T1-T4). Since RMSE values from heterogenous dataset approach were higher than the values in Table 4, these results have not been shown in the present work. The failure of heterogenous dataset approach to capture 2D-3D interface structures accurately is primarily on account of the unique microstructural characteristics of 2D-3D interface from its individual 2D-1D interface counterparts. While the present study emphasizes on a correct description of atomic energies in neural networks, there is further scope to develop successful data sampling approaches for 2D-3D interfaces.

The current state of existing ML models is still miles away from completely replacing DFT for complex interface systems. However, for the purpose of molecular dynamics simulations, modified HDNN models trained with appropriate atomic energy decomposition could be sufficient. Applications of ML-based PES depend on the dataset sampling approach adopted. While AIMD generated dataset trained model could be successfully used for simulations where more than one equilibrium is searched for, DFT based dataset could be sufficient to explore one global minimum for the structures with a certain tradeoff between accuracy and computation times. Ongoing advancements in training and sampling techniques can possibly overcome the challenges associated with ML-assisted interface studies.

## 4. 2D-3D INTERFACES IN ELECTROCHEMICAL ENERGY STORAGE

The conventional bulk materials-based batteries (Fig 6a) have practical issues to meet the ever-increasing energy demand [75]. The two most critical chemo-mechanical failure modes in batteries are – (i) interface failure leading to electrical isolation of active electrode particles [76] and (ii) mechanical failure of active materials [77-79]. The active electrode particles (e.g., Si) have to be in contact with the metal current collector (Fig. 6b1), such as Ni, to ensure a uniform electron exchange [80]. During Li intercalation, active particles undergo substantial volume expansion [81]. For example, Si undergoes 300% volume expansion upon lithiation [82]. Since the metal current collectors (e.g., Cu, Ni) acts as non-slippery surfaces, the volume expansion/contraction of active particles generates excessive interfacial stress (Fig. 6b1), leading to fracture of active particles, causing battery failures [62].

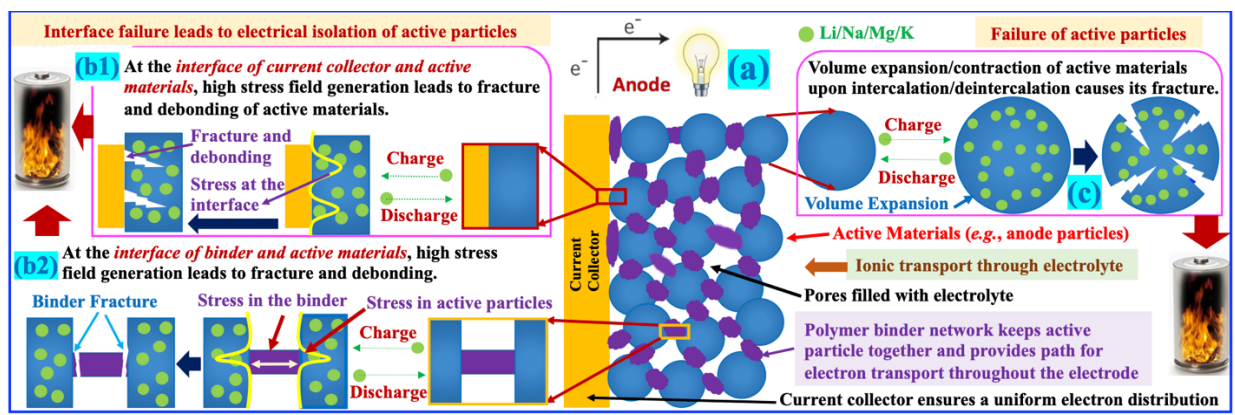

**Figure 6: (a)** Schematic of anode, **(b)** Failure at the interface of binder/active-materials and current-collector/active-materials, **(c)** Failure of active materials.

Another critical interface is between the polymer binder and active particles [83] (Fig. 6b2). The polymer binder network keeps active particles adhered together and ensures continued electrical contact throughout the electrode [84]. However, volume expansion/contraction of active particles causes excessive interfacial stress at polymer-active particle interface, leading to detachment and electrical isolation of active particles [83] (Fig. 6b2). Besides interface failure, another prominent reason for battery failure is the active particles' fracture [85] (Fig. 6c). Again volume expansion/contraction of Si anode particle upon lithiation/delithiation causes its formation of cracks [86], leading to battery failure [87]. These practical problems in battery electrodes need to be addressed by strategizing the electrode design.

2D materials-based heterostructures are promising candidates to solve these burning issues [88-90]. Electrical isolation of active particles can be avoided by replacing polymer binders with conductive and flexible 2D material such as MXenes, which can form an omnipresent electron conducting network within the electrode [91] (Fig. 7a1). Next, the issue of the current collector and active particle interface failure (Fig. 7b1), can be addressed with two options: (i) Fig. 7a2: addition of 2DM such as graphene as 'coating' on the current collector to make it a 'slippery' interface [24, 62], (ii) Fig. 7a3: completely replace the current collector with 2DM such as MXenes [92, 93]. On the other hand, the problem of active materials failure (Fig. 6c) can be solved by using 2D materials-based anode [10] (Fig. 7b1) or 3D active materials integrated with 2D materials (Fig. 7b2)[7, 94].

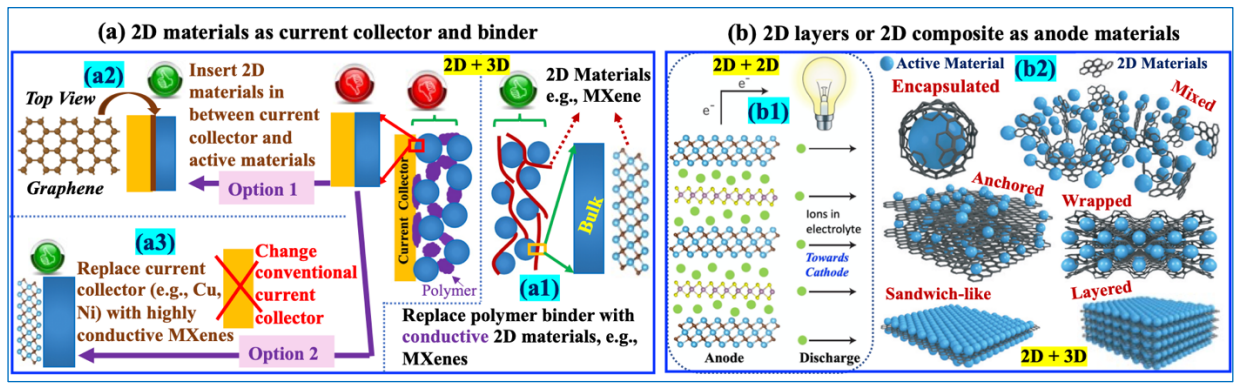

**Figure 7: (a)** 2D materials as current collector and binder, **(b)** 2D layers and 2D composite as anode materials.

In all the 2D materials-based cases discussed, interface plays a critical role in electrochemical performance, i.e., energy and power density, volumetric capacity, etc. The mechanical integrity of these interfaces dictates the long-term performance of energy storage systems [95-97]. Computational modeling methods such as DFT and MD simulations have been good alternatives to expensive experimental characterization for developing a deeper understanding of complex interfacial characteristics in batteries. Work by Basu et al.[98] on recognized benefits of slippery graphene surface at current collector end in combating stresses in Si anode upon lithiation, is an example of comprehensive scope of simulation studies. However,

MD-based methodology to study interfacial stress cannot be extended to new and heavy metal electrodes such as Sn, Se and more, due to lack of appropriate forcefields. Presented work lays the foundation of developing the futuristic AI based potentials to study a wide variety of 2D materials-based interfaces. Although the present study considers only graphene-based interfaces, the presented approach can be implemented in other 2D materials-based systems.

## 5. CONCLUSIONS

In summary, we performed optimization of different graphene and Sn-based 2D-3D interfaces resulting in a unique dataset, a kind lacking in existing databases. Our DFT simulation results show lattice distortions in Sn interfacing with graphene in great atomic details and highlight the preferred stability of β-Sn over graphene as opposed to α-Sn. One of the best application cases of Gr|Sn interface systems is in Sodium ion batteries, where the presence of graphene interface can alleviate mechanical stresses upon Na intercalation in otherwise high capacity-low stability Sn anode. Usage of graphene-based heterostructures is undisputedly vast and the need to model structural-functional aspects of such interfaces is an emergent need. We present the development of ML-based PES that can predict the energies of complex graphene-Sn 2D|3D interface systems with good accuracies that could be used to replace expensive ab-initio methods in the future modeling efforts. Applicability of high dimensional neural networks (HDNN) developed by Behler and Parrinello that utilize atom-centered symmetry functions as structural descriptors has been shown. The widely used approach to calculate loss function on atomic energies showed good performance on validation split but failed to predict energies of the new interface systems. To overcome this, we modify HDNN model to enable training on the total energies of the system rather than atomic energies. This latter approach significantly improves the performance of PES in predicting the total energies of new Gr|Sn interface systems that constitute the test interfaces. Primary reason for this enhanced performance is the freedom model gained to assign atomic energies based on atomic chemical environment. This opens the possibility for more accurate evaluation of atomic energies and forces from ML models, allowing the scope for automated equilibration.

## SUPPORTING INFORMATION

Cut off radius (Rc) for atom-centered symmetry function (ACSF); Initial Structures of Gr|Sn Interfaces in Training Dataset; Equilibration of test interface structures; Structural transformations in interface structures with increased vacuum; Lattice Mismatch in Crystalline Sn over Graphene; Heterogeneous dataset.

## AUTHOR INFORMATION

### Corresponding Authors

Vidushi Sharma is now at IBM Almaden Research Center, San Jose, CA 95120, USA. Email: vs574@njit.edu, vidushis@ibm.com

Dibakar Datta; Email: dibakar.datta@njit.edu

### Author Contributions

V.S. and D.D. conceived the project. V.S. performed all work and wrote the manuscript. Both authors approved the final version of the manuscript.

## CONFLICT OF INTEREST STATEMENT

The authors have no conflicts of interest to declare.

All authors have seen and agree with the contents of the manuscript and there is no financial interest to report. We certify that the submission is original work and is not under review at any other publication.

## ACKNOWLEDGEMENT

The work is supported by National Science Foundation (NSF) Civil, Mechanical and Manufacturing Innovation (CMMI) Program, Award Number # 1911900. Authors acknowledge Extreme Science and Engineering Discovery Environment (XSEDE) for the computational facilities (Award Number – DMR180013). V.S. acknowledges Mr. Joy Datta for fruitful discussion.

## DATA AVAILABILITY

The data reported in this paper is available from the corresponding author upon reasonable request.

## CODE AVAILABILITY

The pre- and post-processing codes used in this paper are available from the corresponding author upon reasonable request. Restrictions apply to the availability of the simulation codes, which were used under license for this study.

## REFERENCES

[1] H. Abdullahi, C.L. Burcham, T. Vetter, A mechanistic model to predict droplet drying history and particle shell formation in multicomponent systems, Chemical Engineering Science 224 (2020) 115713.

[2] K. Chatterjee, S. Sarkar, K.J. Rao, S. Paria, Core/shell nanoparticles in biomedical applications, Advances in colloid and interface science 209 (2014) 8-39.

[3] S. Wei, Q. Wang, J. Zhu, L. Sun, H. Lin, Z. Guo, Multifunctional composite core–shell nanoparticles, Nanoscale 3(11) (2011) 4474-4502.

[4] D. Yan, M. Wei, Photofunctional layered materials, Springer2015.

[5] J. Di, J. Xia, H. Li, S. Guo, S. Dai, Bismuth oxyhalide layered materials for energy and environmental applications, Nano Energy 41 (2017) 172-192.

[6] A. Alanazi, C. Nojiri, T. Kido, j. Noguchi, Y. Ohgoe, T. Matsuda, K. Hirakuri, A. Funakubo, K. Sakai, Y. Fukui, Engineering analysis of diamond-like carbon coated polymeric materials for biomedical applications, Artificial Organs 24(8) (2000) 624-627.

[7] S.-H. Bae, H. Kum, W. Kong, Y. Kim, C. Choi, B. Lee, P. Lin, Y. Park, J. Kim, Integration of bulk materials with two-dimensional materials for physical coupling and applications, Nature materials 18(6) (2019) 550-560.

[8] D.S. Schulman, A.J. Arnold, S. Das, Contact engineering for 2D materials and devices, Chemical Society Reviews 47(9) (2018) 3037-3058.

[9] L. Oakes, R. Carter, T. Hanken, A.P. Cohn, K. Share, B. Schmidt, C.L. Pint, Interface strain in vertically stacked two-dimensional heterostructured carbon-MoS2 nanosheets controls electrochemical reactivity, Nature communications 7(1) (2016) 1-7.

[10] K.-S. Chen, I. Balla, N.S. Luu, M.C. Hersam, Emerging opportunities for two-dimensional materials in lithium-ion batteries, ACS Energy Letters 2(9) (2017) 2026-2034.

[11] Z. Hu, Q. Liu, S.-L. Chou, S.-X. Dou, Two-dimensional material-based heterostructures for rechargeable batteries, Cell Reports Physical Science 2(1) (2021) 100286.

[12] K.S. Novoselov, D. Jiang, F. Schedin, T. Booth, V. Khotkevich, S. Morozov, A.K. Geim, Two-dimensional atomic crystals, Proceedings of the National Academy of Sciences 102(30) (2005) 10451-10453.

[13] H. Hong, C. Liu, T. Cao, C. Jin, S. Wang, F. Wang, K. Liu, Interfacial engineering of van der waals coupled 2D layered materials, Advanced Materials Interfaces 4(9) (2017) 1601054.

[14] Q. Zhang, G. Fiori, G. Iannaccone, On transport in vertical graphene heterostructures, IEEE Electron Device Letters 35(9) (2014) 966-968.

[15] B. Huang, H. Xiang, J. Yu, S.-H. Wei, Effective control of the charge and magnetic states of transition-metal atoms on single-layer boron nitride, Physical review letters 108(20) (2012) 206802.

[16] G. Konstantatos, M. Badioli, L. Gaudreau, J. Osmond, M. Bernechea, F.P.G. De Arquer, F. Gatti, F.H. Koppens, Hybrid graphene–quantum dot phototransistors with ultrahigh gain, Nature nanotechnology 7(6) (2012) 363-368.

[17] Z.Y. Al Balushi, K. Wang, R.K. Ghosh, R.A. Vilá, S.M. Eichfeld, J.D. Caldwell, X. Qin, Y.-C. Lin, P.A. DeSario, G. Stone, Two-dimensional gallium nitride realized via graphene encapsulation, Nature materials 15(11) (2016) 1166-1171.

[18] T. Journot, V. Bouchiat, B. Gayral, J. Dijon, B. Hyot, Self-assembled UV photodetector made by direct epitaxial GaN growth on graphene, ACS applied materials & interfaces 10(22) (2018) 18857-18862.

[19] M. Dutta, S. Sarkar, T. Ghosh, D. Basak, ZnO/graphene quantum dot solid-state solar cell, The Journal of Physical Chemistry C 116(38) (2012) 20127-20131.

[20] S. Sun, L. Gao, Y. Liu, Enhanced dye-sensitized solar cell using graphene-TiO 2 photoanode prepared by heterogeneous coagulation, Applied physics letters 96(8) (2010) 083113.

[21] C.-Y. Chou, G.S. Hwang, Role of interface in the lithiation of silicon-graphene composites: A first principles study, The Journal of Physical Chemistry C 117(19) (2013) 9598-9604.

[22] S.-L. Chou, J.-Z. Wang, M. Choucair, H.-K. Liu, J.A. Stride, S.-X. Dou, Enhanced reversible lithium storage in a nanosize silicon/graphene composite, Electrochemistry Communications 12(2) (2010) 303-306.

[23] Y. Li, S. Huang, C. Wei, D. Zhou, B. Li, C. Wu, V.N. Mochalin, Adhesion Between MXenes and Other 2D Materials, ACS Applied Materials & Interfaces 13(3) (2021) 4682-4691.

[24] V. Sharma, D. Mitlin, D. Datta, Understanding the Strength of the Selenium–Graphene Interfaces for Energy Storage Systems, Langmuir 37(6) (2021) 2029-2039.

[25] P. Khomyakov, G. Giovannetti, P. Rusu, G.v. Brocks, J. Van den Brink, P.J. Kelly, First-principles study of the interaction and charge transfer between graphene and metals, Physical Review B 79(19) (2009) 195425.

[26] J. Behler, Atom-centered symmetry functions for constructing high-dimensional neural network potentials, The Journal of chemical physics 134(7) (2011) 074106.

[27] Y. Zuo, C. Chen, X. Li, Z. Deng, Y. Chen, J.r. Behler, G. Csányi, A.V. Shapeev, A.P. Thompson, M.A. Wood, Performance and cost assessment of machine learning interatomic potentials, The Journal of Physical Chemistry A 124(4) (2020) 731-745.

[28] A.P. Thompson, L.P. Swiler, C.R. Trott, S.M. Foiles, G.J. Tucker, Spectral neighbor analysis method for automated generation of quantum-accurate interatomic potentials, Journal of Computational Physics 285 (2015) 316-330.

[29] A.V. Shapeev, Moment tensor potentials: A class of systematically improvable interatomic potentials, Multiscale Modeling & Simulation 14(3) (2016) 1153-1173.

[30] A.P. Bartók, M.C. Payne, R. Kondor, G. Csányi, Gaussian approximation potentials: The accuracy of quantum mechanics, without the electrons, Physical review letters 104(13) (2010) 136403.

[31] S. Fujikake, V.L. Deringer, T.H. Lee, M. Krynski, S.R. Elliott, G. Csányi, Gaussian approximation potential modeling of lithium intercalation in carbon nanostructures, The Journal of chemical physics 148(24) (2018) 241714.

[32] J. Behler, M. Parrinello, Generalized neural-network representation of high-dimensional potential-energy surfaces, Physical review letters 98(14) (2007) 146401.

[33] P.W. Battaglia, J.B. Hamrick, V. Bapst, A. Sanchez-Gonzalez, V. Zambaldi, M. Malinowski, A. Tacchetti, D. Raposo, A. Santoro, R. Faulkner, Relational inductive biases, deep learning, and graph networks, arXiv preprint arXiv:1806.01261 (2018).

[34] K.T. Schütt, P.-J. Kindermans, H.E. Sauceda, S. Chmiela, A. Tkatchenko, K.-R. Müller, Schnet: A continuous-filter convolutional neural network for modeling quantum interactions, arXiv preprint arXiv:1706.08566 (2017).

[35] H. Yanxon, D. Zagaceta, B.C. Wood, Q. Zhu, Neural network potential from bispectrum components: A case study on crystalline silicon, The Journal of Chemical Physics 153(5) (2020) 054118.

[36] S. Kondati Natarajan, J.r. Behler, Self-Diffusion of surface defects at copper–water interfaces, The Journal of Physical Chemistry C 121(8) (2017) 4368-4383.

[37] N. Artrith, A.M. Kolpak, Understanding the composition and activity of electrocatalytic nanoalloys in aqueous solvents: A combination of DFT and accurate neural network potentials, Nano letters 14(5) (2014) 2670-2676.

[38] J. Behler, R. Martoňák, D. Donadio, M. Parrinello, Metadynamics simulations of the high-pressure phases of silicon employing a high-dimensional neural network potential, Physical review letters 100(18) (2008) 185501.

[39] L. Zhang, J. Han, H. Wang, R. Car, E. Weinan, Deep potential molecular dynamics: a scalable model with the accuracy of quantum mechanics, Physical review letters 120(14) (2018) 143001.

[40] J.S. Smith, O. Isayev, A.E. Roitberg, ANI-1: an extensible neural network potential with DFT accuracy at force field computational cost, Chemical science 8(4) (2017) 3192-3203.

[41] K.T. Schütt, F. Arbabzadah, S. Chmiela, K.R. Müller, A. Tkatchenko, Quantum-chemical insights from deep tensor neural networks, Nature communications 8(1) (2017) 1-8.

[42] K. Yao, J.E. Herr, D.W. Toth, R. Mckintyre, J. Parkhill, The TensorMol-0.1 model chemistry: a neural network augmented with long-range physics, Chemical science 9(8) (2018) 2261-2269.

[43] K. Yao, J.E. Herr, S.N. Brown, J. Parkhill, Intrinsic bond energies from a bonds-in-molecules neural network, The journal of physical chemistry letters 8(12) (2017) 2689-2694.

[44] T. Bereau, D. Andrienko, O.A. Von Lilienfeld, Transferable atomic multipole machine learning models for small organic molecules, Journal of chemical theory and computation 11(7) (2015) 3225-3233.

[45] A.P. Bartók, J. Kermode, N. Bernstein, G. Csányi, Machine learning a general-purpose interatomic potential for silicon, Physical Review X 8(4) (2018) 041048.

[46] V.L. Deringer, N. Bernstein, A.P. Bartók, M.J. Cliffe, R.N. Kerber, L.E. Marbella, C.P. Grey, S.R. Elliott, G. Csányi, Realistic atomistic structure of amorphous silicon from machine-learning-driven molecular dynamics, The journal of physical chemistry letters 9(11) (2018) 2879-2885.

[47] N. Xu, Y. Shi, Y. He, Q. Shao, A Deep-Learning Potential for Crystalline and Amorphous Li–Si Alloys, The Journal of Physical Chemistry C 124(30) (2020) 16278-16288.

[48] C.M. Andolina, P. Williamson, W.A. Saidi, Optimization and validation of a deep learning CuZr atomistic potential: Robust applications for crystalline and amorphous phases with near-DFT accuracy, The Journal of chemical physics 152(15) (2020) 154701.

[49] L. Tang, Z. Yang, T. Wen, K.-M. Ho, M.J. Kramer, C.-Z. Wang, Development of interatomic potential for Al–Tb alloys using a deep neural network learning method, Physical Chemistry Chemical Physics 22(33) (2020) 18467-18479.

[50] H.R. Banjade, S. Hauri, S. Zhang, F. Ricci, W. Gong, G. Hautier, S. Vucetic, Q. Yan, Structure motif–centric learning framework for inorganic crystalline systems, Science Advances 7(17) (2021) eabf1754.

[51] N.C. Frey, D. Akinwande, D. Jariwala, V.B. Shenoy, Machine Learning-Enabled Design of Point Defects in 2D Materials for Quantum and Neuromorphic Information Processing, ACS nano 14(10) (2020) 13406-13417.

[52] K. Tanaka, K. Hachiya, W. Zhang, K. Matsuda, Y. Miyauchi, Machine-Learning Analysis to Predict the Exciton Valley Polarization Landscape of 2D Semiconductors, ACS nano 13(11) (2019) 12687-12693.

[53] Y.J. jae Shin, W. Shin, T. Taniguchi, K. Watanabe, P. Kim, S.-H. Bae, Fast and accurate robotic optical detection of exfoliated graphene and hexagonal boron nitride by deep neural networks, 2D Materials  (2020).

[54] M. Fernández, S. Rezaei, J.R. Mianroodi, F. Fritzen, S. Reese, Application of artificial neural networks for the prediction of interface mechanics: a study on grain boundary constitutive behavior, Advanced Modeling and Simulation in Engineering Sciences 7(1) (2020) 1-27.

[55] J.r. Behler, Four Generations of High-Dimensional Neural Network Potentials, Chemical Reviews  (2021).

[56] J. Han, L. Zhang, R. Car, Deep potential: A general representation of a many-body potential energy surface, arXiv preprint arXiv:1707.01478  (2017).

[57] M. Gastegger, L. Schwiedrzik, M. Bittermann, F. Berzsenyi, P. Marquetand, wACSF—Weighted atom-centered symmetry functions as descriptors in machine learning potentials, The Journal of chemical physics 148(24) (2018) 241709.

[58] L. Himanen, M.O. Jäger, E.V. Morooka, F.F. Canova, Y.S. Ranawat, D.Z. Gao, P. Rinke, A.S. Foster, DScribe: Library of descriptors for machine learning in materials science, Computer Physics Communications 247 (2020) 106949.

[59] H. Gao, J. Wang, J. Sun, Improve the performance of machine-learning potentials by optimizing descriptors, The Journal of chemical physics 150(24) (2019) 244110.

[60] H. Yanxon, D. Zagaceta, B. Tang, D.S. Matteson, Q. Zhu, PyXtal_FF: a python library for automated force field generation, Machine Learning: Science and Technology 2(2) (2020) 027001.

[61] G. Ceder, K. Persson, The materials project: A materials genome approach, 2010.

[62] S. Basu, S. Suresh, K. Ghatak, S.F. Bartolucci, T. Gupta, P. Hundekar, R. Kumar, T.-M. Lu, D. Datta, Y. Shi, Utilizing van der Waals slippery interfaces to enhance the electrochemical stability of silicon film anodes in lithium-ion batteries, ACS applied materials & interfaces 10(16) (2018) 13442-13451.

[63] V. Sharma, K. Ghatak, D. Datta, Amorphous germanium as a promising anode material for sodium ion batteries: a first principle study, Journal of Materials Science 53(20) (2018) 14423-14434.

[64] G. Kresse, J. Furthmüller, Efficient iterative schemes for ab initio total-energy calculations using a plane-wave basis set, Physical review B 54(16) (1996) 11169.

[65] G. Kresse, D. Joubert, From ultrasoft pseudopotentials to the projector augmented-wave method, Physical Review B 59(3) (1999) 1758.

[66] P.E. Blöchl, Projector augmented-wave method, Physical review B 50(24) (1994) 17953.

[67] J.P. Perdew, K. Burke, M. Ernzerhof, Generalized gradient approximation made simple, Physical review letters 77(18) (1996) 3865.

[68] M. Dion, H. Rydberg, E. Schröder, D.C. Langreth, B.I. Lundqvist, Van der Waals density functional for general geometries, Physical review letters 92(24) (2004) 246401.

[69] F. Legrain, S. Manzhos, Understanding the difference in cohesive energies between alpha and beta tin in DFT calculations, AIP Advances 6(4) (2016) 045116.

[70] B. Luo, T. Qiu, D. Ye, L. Wang, L. Zhi, Tin nanoparticles encapsulated in graphene backboned carbonaceous foams as high-performance anodes for lithium-ion and sodium-ion storage, Nano Energy 22 (2016) 232-240.

[71] E. Sanville, S.D. Kenny, R. Smith, G. Henkelman, Improved grid-based algorithm for Bader charge allocation, Journal of computational chemistry 28(5) (2007) 899-908.

[72] Z. Zhang, Improved adam optimizer for deep neural networks, 2018 IEEE/ACM 26th International Symposium on Quality of Service (IWQoS), IEEE, 2018, pp. 1-2.

[73] M. Comin, L.J. Lewis, Deep-learning approach to the structure of amorphous silicon, Physical Review B 100(9) (2019) 094107.

[74] S.-D. Huang, C. Shang, P.-L. Kang, Z.-P. Liu, Atomic structure of boron resolved using machine learning and global sampling, Chemical science 9(46) (2018) 8644-8655.

[75] S. Chu, A. Majumdar, Opportunities and challenges for a sustainable energy future, nature 488(7411) (2012) 294-303.

[76] B. Liang, Y. Liu, Y. Xu, Silicon-based materials as high capacity anodes for next generation lithium ion batteries, Journal of Power sources 267 (2014) 469-490.

[77] X.H. Liu, L. Zhong, S. Huang, S.X. Mao, T. Zhu, J.Y. Huang, Size-dependent fracture of silicon nanoparticles during lithiation, ACS nano 6(2) (2012) 1522-1531.

[78] Y. Jin, B. Zhu, Z. Lu, N. Liu, J. Zhu, Challenges and recent progress in the development of Si anodes for lithium-ion battery, Advanced Energy Materials 7(23) (2017) 1700715.

[79] S.W. Lee, I. Ryu, W.D. Nix, Y. Cui, Fracture of crystalline germanium during electrochemical lithium insertion, Extreme Mechanics Letters 2 (2015) 15-19.

[80] L. Zhang, H. Gong, Partial conversion of current collectors into nickel copper oxide electrode materials for high-performance energy storage devices, ACS applied materials & interfaces 7(28) (2015) 15277-15284.

[81] B. Jerliu, E. Hüger, L. Dorrer, B.-K. Seidlhofer, R. Steitz, V. Oberst, U. Geckle, M. Bruns, H. Schmidt, Volume expansion during lithiation of amorphous silicon thin film electrodes studied by

in-operando neutron reflectometry, The Journal of Physical Chemistry C 118(18) (2014) 9395-9399.

[82] M. Ko, S. Chae, J. Cho, Challenges in accommodating volume change of Si anodes for Li-ion batteries, ChemElectroChem 2(11) (2015) 1645-1651.

[83] A. Santimetaneedol, R. Tripuraneni, S.A. Chester, S.P. Nadimpalli, Time-dependent deformation behavior of polyvinylidene fluoride binder: Implications on the mechanics of composite electrodes, Journal of Power Sources 332 (2016) 118-128.

[84] W. Zeng, L. Wang, X. Peng, T. Liu, Y. Jiang, F. Qin, L. Hu, P.K. Chu, K. Huo, Y. Zhou, Enhanced ion conductivity in conducting polymer binder for high-performance silicon anodes in advanced lithium-ion batteries, Advanced Energy Materials 8(11) (2018) 1702314.

[85] S. Kalnaus, K. Rhodes, C. Daniel, A study of lithium ion intercalation induced fracture of silicon particles used as anode material in Li-ion battery, Journal of Power Sources 196(19) (2011) 8116-8124.

[86] F. Fan, S. Huang, H. Yang, M. Raju, D. Datta, V.B. Shenoy, A.C. Van Duin, S. Zhang, T. Zhu, Mechanical properties of amorphous LixSi alloys: a reactive force field study, Modelling and Simulation in Materials Science and Engineering 21(7) (2013) 074002.

[87] S.W. Lee, M.T. McDowell, L.A. Berla, W.D. Nix, Y. Cui, Fracture of crystalline silicon nanopillars during electrochemical lithium insertion, Proceedings of the National Academy of Sciences 109(11) (2012) 4080-4085.

[88] L. Shi, T. Zhao, Recent advances in inorganic 2D materials and their applications in lithium and sodium batteries, Journal of Materials Chemistry A 5(8) (2017) 3735-3758.

[89] Y. Dong, Z.-S. Wu, W. Ren, H.-M. Cheng, X. Bao, Graphene: a promising 2D material for electrochemical energy storage, Science Bulletin 62(10) (2017) 724-740.

[90] E. Pomerantseva, Y. Gogotsi, Two-dimensional heterostructures for energy storage, Nature Energy 2(7) (2017) 1-6.

[91] C.J. Zhang, S.-H. Park, A. Seral-Ascaso, S. Barwich, N. McEvoy, C.S. Boland, J.N. Coleman, Y. Gogotsi, V. Nicolosi, High capacity silicon anodes enabled by MXene viscous aqueous ink, Nature communications 10(1) (2019) 1-9.

[92] C.-H. Wang, N. Kurra, M. Alhabeb, J.-K. Chang, H.N. Alshareef, Y. Gogotsi, Titanium carbide (MXene) as a current collector for lithium-ion batteries, ACS omega 3(10) (2018) 12489-12494.

[93] V. Sharma, D. Datta, Variation in the interface strength of silicon with surface engineered Ti 3 C 2 MXenes, Physical Chemistry Chemical Physics 23(9) (2021) 5540-5550.

[94] R. Raccichini, A. Varzi, S. Passerini, B. Scrosati, The role of graphene for electrochemical energy storage, Nature materials 14(3) (2015) 271-279.

[95] K. Zhao, Y. Cui, Understanding the role of mechanics in energy materials: A perspective, Extreme Mechanics Letters 9 (2016) 347-352.

[96] R.M. McMeeking, R. Purkayastha, The role of solid mechanics in electrochemical energy systems such as lithium-ion batteries, Procedia IUTAM 10 (2014) 294-306.

[97] S.Z. Butler, S.M. Hollen, L. Cao, Y. Cui, J.A. Gupta, H.R. Gutiérrez, T.F. Heinz, S.S. Hong, J. Huang, A.F. Ismach, Progress, challenges, and opportunities in two-dimensional materials beyond graphene, ACS nano 7(4) (2013) 2898-2926.

[98] S. Basu, S. Suresh, K. Ghatak, S.F. Bartolucci, T. Gupta, P. Hundekar, R. Kumar, T.-M. Lu, D. Datta, Y. Shi, Utilizing van der Waals slippery interfaces to enhance the electrochemical stability of Silicon film anodes in Lithium-ion batteries, ACS applied materials & interfaces (2018).

S 1